\documentclass[manuscript]{aastex}

\newcommand{\myemail}{rmennick@astro-udec.cl}

\slugcomment{Submitted to PASP.}

\shorttitle{L-Band Spectra of 13 Outbursting Be Stars}
\shortauthors{Mennickent et al.}

\begin{document}


\title{L-Band Spectra of 13 Outbursting Be Stars}


\author{R. Mennickent\altaffilmark{1}}
\affil{Departamento de F\'{\i}sica, Facultad de Ciencias F\'{\i}sicas y
Matem\'aticas, Universidad de Concepci\'on, Casilla 160-C, Concepci\'on, Chile}
\email{\myemail}

\author{B. Sabogal}
\affil{Departamento de F\'{\i}sica,   Universidad de Los Andes, Carrera 1 No. 18A-10, Edificio H, Bogot\'a, Colombia}

\and

\author{A. Granada\altaffilmark{2} and L. Cidale\altaffilmark{3}}
\affil{Facultad de Ciencias Astron\'omicas y Geof\'isicas,  
Universidad Nacional de La Plata and Instituto de Astrofisica La Plata (IALP-CCT), \\CONICET, Paseo del Bosque S/N, 1900, La Plata, Argentina}


\altaffiltext{1}{Based on ESO proposal 71.C-0317(B)}
\altaffiltext{2}{Fellow of CONICET}
\altaffiltext{3}{Member of the Carrera del Investigador, CONICET}


\begin{abstract}
We present new L-band spectra of 13 outbursting Be stars obtained with ISAAC at the ESO Paranal observatory. 
These stars can be classified in three groups depending on the presence/absence of emission lines and the strength of  Br$\alpha$ and Pf$\gamma$ emission lines relative to those of Humphreys lines
from transitions 6--14 to the end of the series. These groups are representative of  circumstellar envelopes with different optical depths. 
For the group showing   Br$\alpha$ and Pf$\gamma$ lines stronger than Humphreys lines,   the Humphreys decrement roughly follow the Menzel case-B  for optically thin conditions. For the group showing comparable Br$\alpha$, Pf$\gamma$ and Humphreys emission line strengths, the Humphreys decrements moves from an optically thin to an optically thick regime at a transition wavelength which is characteristic for each star, but typically is located around 3.65--3.75 $\mu$m
(transitions 6--19 and 6--17). Higher order Humphreys lines probe optically thin inner regions even in the optically thicker envelopes.
We find evidence of larger broadening in the infrared emission lines compared with optical lines, 
probably reflecting larger vertical velocity fields near the star.   
The existence of the aforementioned groups is in principle consistent with the proposed description by de Wit et al. (2006) for Be star outbursts in terms of the ejection of an optically thick disk that expands and becomes optically thin before dissipation into the interstellar medium. Time resolved L-band 
spectroscopy sampling the outburst cycle promises to be an unique tool for testing Be star disk evolution.

\end{abstract}





\section{Introduction}

Be stars are rapidly rotating dwarf or giant B type objects that show or have once shown  emissions in the H$\alpha$ line \citep{b17}. Classical Be stars have moderate infrared excesses that originate in the free-free and free-bound emission from ionized circumstellar gas \citep{b12, b45}. Interferometric studies have shown that this gas is concentrated towards the equatorial plane forming a dense disk-like envelope extending extending up to $\approx$ 10 stellar radii  from the stellar surface (Grundstrom 
\& Gies 2006, Quirrenbach et al. 1994, Stee et al. 1995). The  IR region is dominated by broad and bright emission lines arising from high levels of hydrogen atoms \citep{b4,b1,bla}. He\,I, Mg\,II and Na\,I emission lines in the K-band have also been reported (Clark and Steele 2000). The IR line optical depths and line flux ratios display a large variation from star to star \citep{b30} and do not correlate with the spectral type \citep{bla}.  
Be stars are also intrinsically variable, some of them
show mild periodic or irregular photometric variability (Mennickent, Vogt \& Sterken 1994, Sterken, Vogt \& Mennickent 1996), whereas others show sudden brightenings usually attributed to 
mass ejections from the surface of the stars, which can occur discretely over a range of timescales (Hubert, Floquet \& Zorec 2000, Mennickent et al. 2002, de Wit et al. 2006). These outbursts probably induce variability in the opacity, the size and the geometry of the circumstellar envelope. Thus our aim is to explore the  physical properties of envelopes of outbursting Be stars.  To our purpose, we selected for an infrared spectroscopic study 6 Galactic Be stars showing long-lived outbursts (duration several hundred days) and 7 showing short-lived outbursts (duration days or tens of days)  from the list of \cite{b16}. The stars were selected spanning a wide range of projected rotational velocities and most of them have been rarely studied spectroscopically. We hope contributing to the knowledge of the L-band spectral region in Be stars and its relation with the circumstellar envelope,   which have been hitherto scarcely studied.


\section{Observations and data reduction}
\label{Secobs}

\begin{table}
\begin{center}
\tabcolsep 1.5pt
\caption{ List of Be stars
for which L-band spectra were obtained with ISAAC. A quality flag (Q)
is given for every spectrum: G for good and N for noisy} 
\label{obstab}
\tiny
\begin{tabular}{lrcccccrr}
\\
\tableline
Object &         N &   slit & exptime &  airmass & seeing   &     MJD &&    Q\\
       &           &   [\arcsec]& [s] & &  [\arcsec] &         &         & \\
\tableline
\tableline
V1448 Aql  & 20  &   0.3 &  1.85 & 1.3 &  0.6 &   52789.3411-52789.3726&& G\\
V1448 Aql  &  6  &   2.0 &  0.28 & 1.2 &  0.7 &   52789.2966-52789.3046&& N\\
V518 Car   &  4  &   0.3 &  1.85 & 1.5 &  0.7 &   52789.0671-52789.0914&& G\\
V518 Car   &  2 &    2.0 &  0.28 & 1.7 &  0.8 &   52789.1062-52789.1078&& N\\
V767 Cen   &  2 &    2.0 &  0.28 & 1.7 &  0.7 &   52860.0731-52860.0746&& G\\
V817 Cen   &  2 &    0.3 &  1.85 & 1.7 &  0.8 &   52847.0313-52847.0329&& G\\
V817 Cen   &  2 &    2.0 &  0.28 & 2.0 &  0.9 &   52847.0525-52847.0540&& G\\
$\mu$ Cen  &  2 &    2.0 &  0.28 & 1.7 &  1.0 &   52859.0684-52859.0700&& G\\
$\mu$ Cen  &  2 &    0.3 &  0.28 & 1.6 &  0.9 &   52859.0622-52859.0639&& G\\
CP Cir     &  4 &    0.3 &  1.85 & 1.6 &  0.7 &   52788.2476-52788.2526&& G\\
CP Cir     &  2 &    2.0 &  0.28 & 1.4 &  0.8 &   52788.2127-52788.2303&& N\\
CP Cir     &  2 &    2.0 &  0.28 & 1.6 &  0.6 &   52788.2643-52788.2659&& N\\
KV Mus     & 20 &    0.3 &  1.85 & 1.9 & 0.7-1.0 &52858.9951-52859.0246&& N\\
OZ Nor     &  2 &    2.0 &  0.28 & 1.1 &  0.9 &   52789.2050-52789.2066&& G\\
OZ Nor     &  2 &    0.3 &  1.85 & 1.8 &  0.9 &   52790.3767-52790.3784&& G\\
V457 Sct   & 20 &    0.3 &  1.85 & 1.1 &  0.7 &   52788.3288-52788.3603&& G\\
V457 Sct   &  8 &    2.0 &  1.85 & 1.2 &  1.1 &   52789.2186-52789.2297&& N\\
V341 Sge   &  4 &    0.3 &  1.85 & 1.5 &  0.4 &   52789.3885-52789.3918&& G\\
V341 Sge   &  2 &    2.0 &  0.28 & 1.4 &  0.7 &   52789.3138-52789.3155&& G\\
V4024 Sgr  &  2 &    2.0 &  0.28 & 1.0 &  0.7 &   52789.2791-52789.2806&& G\\
V4024 Sgr  &  2 &    0.3 &  1.85 & 1.8 &  0.8 &   52790.3921-52790.3937&& G\\
V1150 Tau  & 16 &    0.3 &  1.85 & 1.2 & 1.0-1.4 &52886.4000-52886.4232&& G\\
V1150 Tau  &  6 &    2.0 &  0.28 & 1.5 &  0.8 &   52887.3249-52887.3324&& G\\
V395 Vul   &  2 &    2.0 &  0.28 & 1.5 &  0.7 &   52789.3258-52789.3274&& G\\
V395 Vul   &  2 &    0.3 &  1.85 & 1.7 &  0.8 &   52790.4069-52790.4086&& G\\
\tableline
\end{tabular}
\end{center}
\end{table}
\normalsize

L-band spectra were obtained with the VLT
Infrared Spectrometer and Array Camera (ISAAC, Moorwood et al. 1998)  at the ESO Cerro Paranal Observatory in service mode during the nights of may 28--30, july 26, august 7--8 and september 3--4, 2003. 
The mode long slit low resolution spectroscopy was selected along with a central wavelength of 3.5 $\mu$m. 
The pixel scale was 0.146 \arcsec/pixel. Two different setups were used, a narrow slit of 0.3\arcsec~ and
other of 2\arcsec, providing resolving power of 1200 and 180, respectively. As our targets are bright, we had to use
very short exposure times (0.3--2s) so that the detector does not become
saturate; this had no effect on the total number of counts recorded, since the exposure time is controlled by the detector and no shutter geometrical bias is involved during the photon acquisition.
Images were reduced with the ISAAC pipeline. The spectra were telluric corrected with the aid of  early
G-type telluric standards observed during the run at similar airmasses that science objects using the procedure 
described in Maiolino, Rieke \& Rieke (1996).
We built our telluric templates by dividing the telluric spectra by a synthetic solar-type spectrum interpolated at the same resolution
and wavelength range. 
Then, we used the IRAF\footnote{IRAF is distributed by the National Optical Astronomy Observatories,
    which are operated by the Association of Universities for Research
    in Astronomy, Inc., under cooperative agreement with the National
    Science Foundation.} telluric task to remove telluric absorption lines from the science objects.
The telluric bands were successfully removed from our spectra, except in some cases in the wavelengths short-ward of 3.4 $\mu$m, characterized by heavy and variable atmospheric absorption. Nevertheless, it has minor importance since the spectral lines used in this work are mostly located out of this region. 
The spectra taken with the wide slit were flux calibrated with the aid of  the
standard star BS5471 (spectral type B3V), whose L magnitude is known. Those spectra taken 
with the narrow slit were continuum normalized. 
The observing log given in Table 1 indicates the number of spectra per object, single spectra exposure times and some additional observational parameters.

\section{Results}
 
The  L-band IR  spectral region, between 3.0 $\mu$m and 4.1 $\mu$m, displays Br$\alpha$, Pf$\gamma$, Pf$\delta$, Pf$\epsilon$ and most of Humphreys' lines from transition 6$-$14 to the end of the series. 
Weak He\,I 4.038-4.041 $\mu$m emission is observed in $\mu$ Cen, OZ Nor,  V\,4024  Sgr and V\,1150 Tau. Although the blue region was discarted by not providing a confident telluric substraction, we do observe Pf$\delta$ and Pf$\epsilon$ emission in OZ Car, V\,817 Cen and V\,1150 Tau.  
The emission line spectra look similar to those of  Be stars with irregular photometric variability, like $\gamma$ Cas (Hony et al. 2000).
According to the intensity of the emission lines, the sample of 13 Be stars  can be classified in three groups: 
those with Br$\alpha$ and Pf$\gamma$ emissions equally intense as Humphreys' lines (Group I, gathered in Fig.\,1); those with
Br$\alpha$ and Pf$\gamma$ emissions more intense than Humphreys' lines (Group II, see Fig.\,2), and those where no line emission is detected (Group III, Fig.\,3).  In this scheme, V\,1150 Tau could be a transition object between Group I and II.
All stars show blue continua, and $\mu$ Cen is notable for showing a break in the continuum slope around $\lambda$ = 3.9 $\mu$m, which we attribute to 
an instrumental effect that should not affect our measurements of line intensity relative to the continuum and emission line widths. 

\begin{figure*}
\epsscale{1}
\plotone{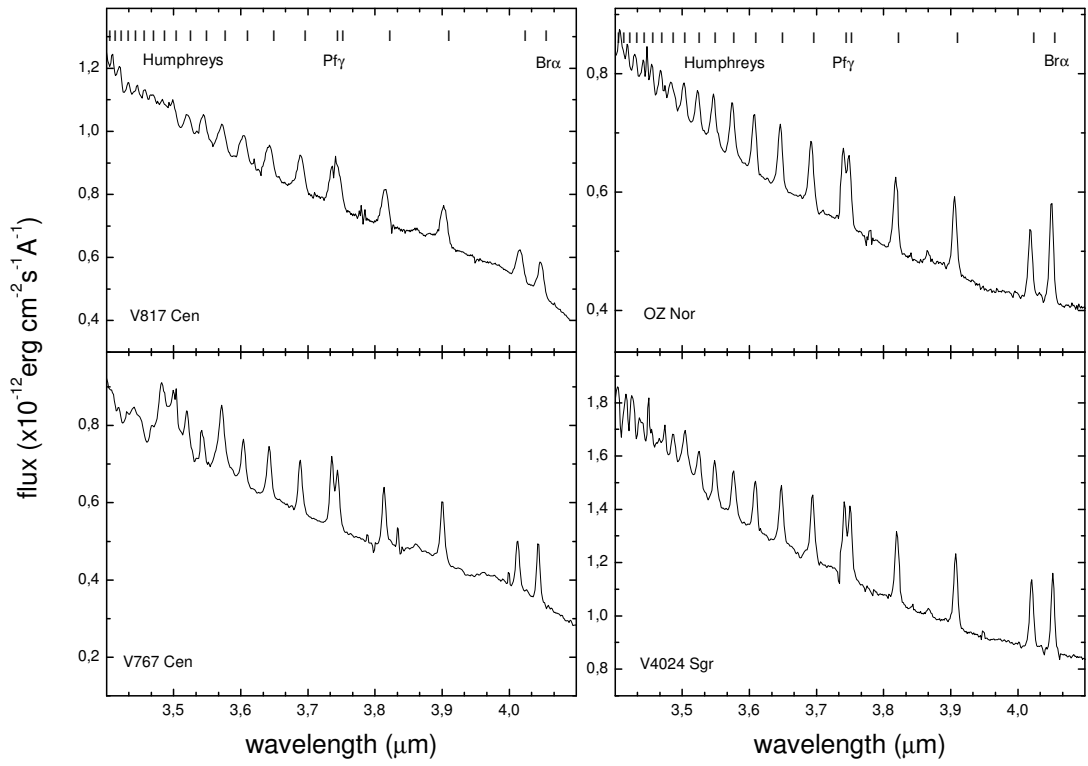}
\caption{IR wide slit spectra of stars of Group I.\label{fig1}}
\end{figure*}

\begin{figure*}
\epsscale{1}
\plotone{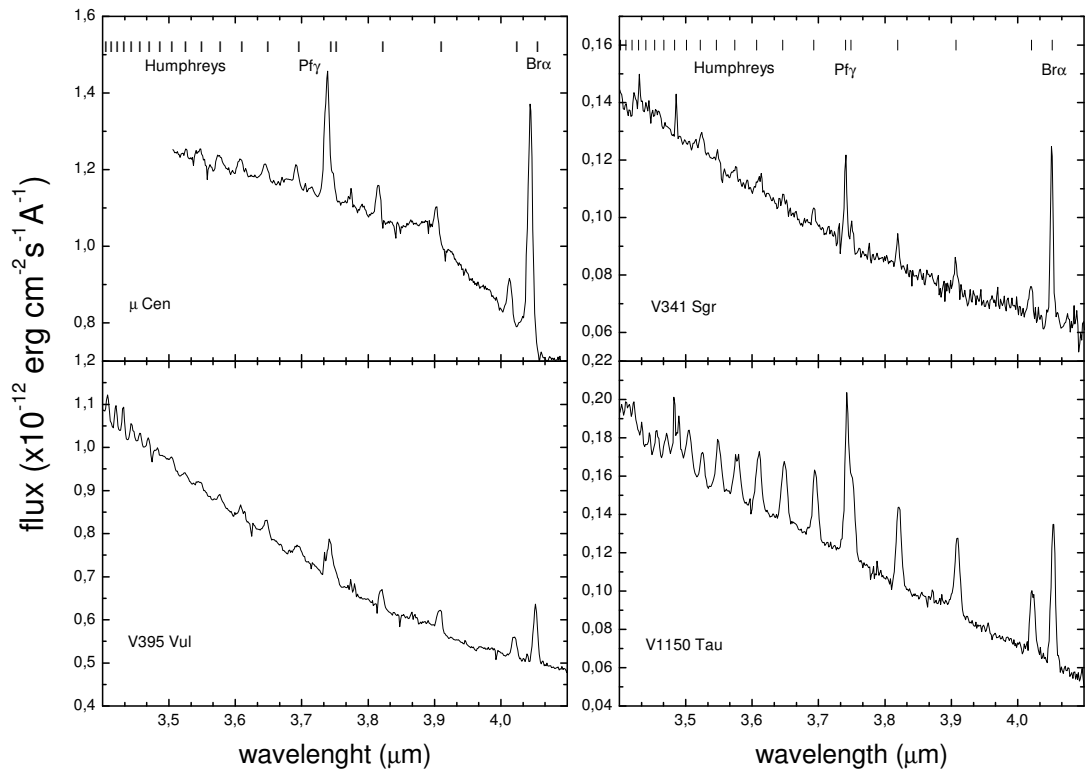}
\caption{IR wide slit spectra of stars of Group II and V\,1150 Tau.\label{fig2}}
\end{figure*}

\begin{figure*}
\epsscale{1}
\plotone{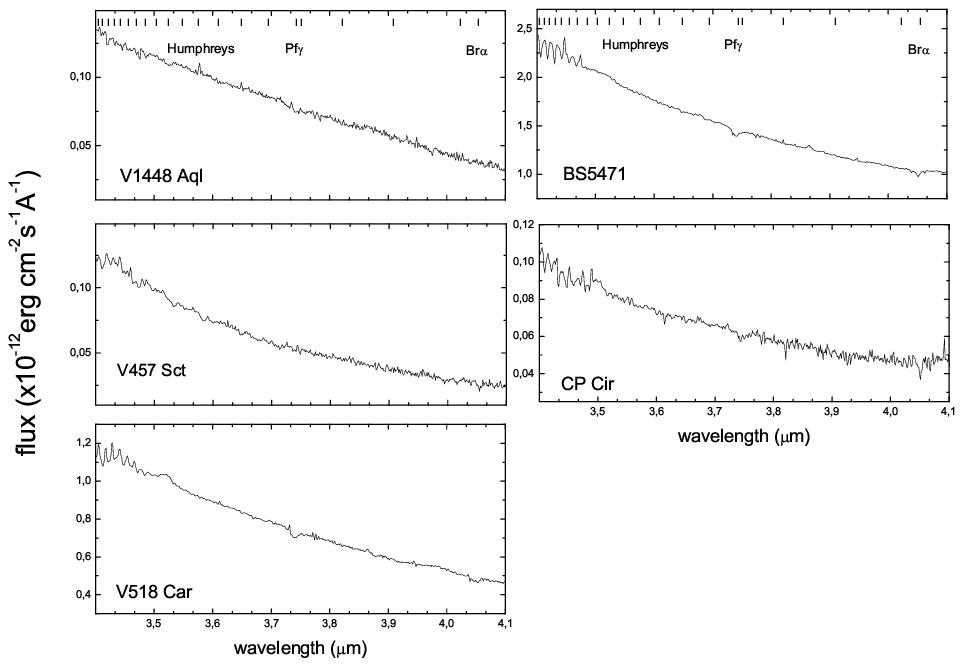}
\caption{IR wide slit spectra of stars of Group III and the B3V star BS5471.\label{fig3}}
\end{figure*}

 L magnitudes were computed by convolution of the spectra obtained with the slit of 2\arcsec~ with the transmission curve of the L filter 
given by \cite{b3}. As part of the spectrum of $\mu$ Cen was
corrupted and for KV Mus we had only one spectrum taken with the narrow slit (0.3\arcsec)  which shows no emission lines, 
no magnitude determinations were possible for these stars.   In Table 2 we list the L magnitudes, the fundamental parameters of the program stars and their classification in the aforementioned groups.
The $V\,sin\,i$ values were taken from \citet{b16}, \citet{b13}, \citet{b48} and  \cite{b11}.

Some of our stars were too bright to determine the outburst stage (outburst/quiescence)  from 
reliable ASAS-3 V-band light curves \citep{p1} at the epoch of our L-band spectroscopy. Others were not included in such a catalog. 
The exception was OZ Nor, that was observed near maximum at the time of the L band spectroscopy. 
The H$\alpha$ spectrum of V\,395 Vul taken in may 23, 2003 and of V\,4024 Sgr taken in may 29, 2003, suggest that our IR observations for these stars were obtained during minimum and maximum H$\alpha$ emission, respectively
(http://astrosurf.com/buil/becat/).

Line fluxes, full width at half maximum ($FWHM$) and equivalent widths, measured in the narrow slit spectra using the splot IRAF task, vary strongly from star to star. In Table 3 we list these parameters for the stars showing emission lines (groups I and II).


\clearpage
\begin{table}
\begin{center}
\caption{Spectrophotometric L magnitudes, fundamental parameters from given references and 2MASS color excess of the observed Be stars. The note indicates the classification group and the outburst character (l= long, s=short, see text). $V\,sin\,i$ is in km\,s$^{-1}$.}
\label{partab}
\tabcolsep 2.5pt
\begin{tabular}{lccccccccccc}
\\
\tableline
~~~object  & L    &T$_{\mathrm{eff}} ($K$)$ & $\log\,g$ && ~~S.T. & $V\,sin\,i$ &$E(H-K)$ & Note  &Ref.\\
\tableline
\tableline
V1448 Aql   & 7.66 &20\,000            & 3.80     & & B2\,IVe  & 243&0.124 & III/s &a\\
V518 Car  & 5.31 &18\,700            & --       & & B3-5\,Ve & 130  &0.043 & III/l  &c\\
$\mu$ Cen   &  --  &23\,130            & 4.04     & & B2\,IV-Ve& 155  &-- & II/s  & d \\
V767 Cen  & 5.66 &23\,320            & 3.95     & & B3\,IIIe &  70  &0.224 & I/s &d\\
V817 Cen   & 5.25 &17\,490            & 3.12     & & B3\,IVe  & 130 &0.178 & I/s &b\\
CP Cir     & 8.00 &15\,000            & --       & & B5\,IVe  & --   &0.005 & III/l  &  c\\
KV Mus     &  --  &18\,700            & --       & & B3\,Ve   & --  &0.057 & III/l  &c\\
OZ Nor    & 5.64 &20\,300            & --       & & B2\,Ve  & 86   &0.259 & I/l &c\\
V457 Sct   & 7.82 &23\,000            & 3.80     & & B1.5\,IVe&202   &0.306 & III/s  &a\\
V341 Sge  & 7.58 &22\,000            & --       & & B2.5\,Ve &115  &0.105 & II/s & c\\
V4024 Sgr   & 4.85 &18\,940            & 3.49     & & B2\,Ve   &120 &0.113 & I/l &d\\
V1150 Tau  & 7.25 &22\,000            & --       & & B2.5\,Ve &340  &0.323  & I/l &c\\
V395 Vul    & 5.36 &17\,790            & 3.93     & & B2.5\,Ve &230  &0.134 & II/s  &d\\
\tableline
\multicolumn{8}{l}{~~}\\
\multicolumn{8}{l}{(a) and (b) from Fr\'emat et al. (2005,2006)}\\
\multicolumn{8}{l}{(c) from Wright et al. (2003);(d) from Zorec et al. (2005)}\\
\end{tabular}
\end{center}
\end{table}

\clearpage
\begin{table*}
\begin{center}
\caption{Spectroscopic measurement. We give the observed line fluxes $f$ ($\pm$ 10\%, in ergs cm$^{-2}$s$^{-1}$), equivalent widths $EW$  ($\pm$ 10\%, in \AA) and full width at half maximum $FWHM$ ($\pm$ 40 km\,s$^{-1}$). $\lambda_{0}$ defined in the text is in microns.}
\label{partab}
\tiny
\tabcolsep 2.5pt
\begin{tabular}{lcccccccccccc}
\\
\tableline
~~~object  &$f_{Br\alpha}$  &   $f_{Pf\gamma}$  &   $f_{Hu14}$  & $ -EW_{ Br\alpha}$ & $-EW_{Pf\gamma}$  &  $ -EW_{Hu14}$&$FWHM_{Br\alpha}$ &        $FWHM_{Pf\gamma}$    &$FWHM_{Hu14}$ &$\lambda_{0}$\\
\tableline
\tableline
$\mu$ Cen  & 5.46E-11 &2.32E-11 &1.03E-11 & 82.8 & 25.8 &  16.0      &512  &370      &447 &$\geq 4.0$  \\
V767 Cen  & 9.73E-12    &1.17E-11   &9.25E-12    &28.3   &21.5   &24.7  &--        &-  -        &--   &3.65    \\
V817 Cen  &1.62E-11 &1.26E-11 &1.23E-11 & 24.9 & 17.2 & 22.7     &277   &399      &357  &3.65\\
OZ Nor   & 1.28E-11 & 9.50E-12 &9.38E-12 & 35.0 & 16.5 & 23.3      &238    &288     & 267  &3.75\\
V341 Sge  & 3.41E-12 & 1.47E-12 & 6.80E-13 & 57.0 & 17.7 &  11.6   &136    &145    &128  &$\geq 4.0$\\
V4024 Sgr    &1.99E-11& 2.23E-11 &1.84E-11 & 23.3 & 15.9 & 19.8   &356   &327     &325   &3.75\\
V1150 Tau &5.91E-12  &5.28E-12 &4.04E-12 & 72.6 & 44.2 & 40.0      &350   &556       &567  &3.75\\
V395 Vul    &1.06E-11 &7.88E-12 &5.58E-12 & 24.3 & 14.3 & 9.9     &538   &554     &484 &$\geq 4.0$\\
\tableline
\end{tabular}
\end{center}
\end{table*}

 \clearpage
\begin{table}
\begin{center}
\caption{Results of the $FWHM-V\,sin\,i$ fits. $R$ is the correlation coefficient and $N$ is the number of stars.}
\label{fwhmtab}
\tabcolsep 2.5pt
\begin{tabular}{lccccccc}
\\
\tableline
Line & A (km/s)  &B &R &$\sigma$ (km/s) &N &rejected&source\\
\tableline
\tableline
Pf$\gamma$ &1.12$\pm$0.25 & 218$\pm$49 &0.83 &52& 6 &1 &a\\
Hu14 &1.12$\pm$0.20 & 209$\pm$39&0.89&41&6&1 &a\\
Hu15&1.62$\pm$0.55 & 163$\pm$103&0.63&120&7&0&a\ \\
Hu16 &2.04$\pm$0.33&48$\pm$61&0.89&71&7&0&a\\
Hu21 &0.68$\pm$0.09&363$\pm$18&0.95&19&5&0&a\\
H$\alpha$ & 1.23 & 70 &- &-& 115 &0 &b\\
H$\beta$+Fe\,II &1.2  & 30 &- &- & 90 &0 &b\\
Br$\gamma$ &0.759  & 149 &- &- & 39 &0 &c\\
He\,I 2.058 &0.643  & 247 &- &- & 19 &0 &c\\
\tableline
\multicolumn{8}{l}{(a) this paper, (b) Hanuschik 1989, (c) Clarke \& Steele 1999}\\
\end{tabular}
\end{center}
\end{table}



\begin{figure}
\epsscale{1}
\plotone{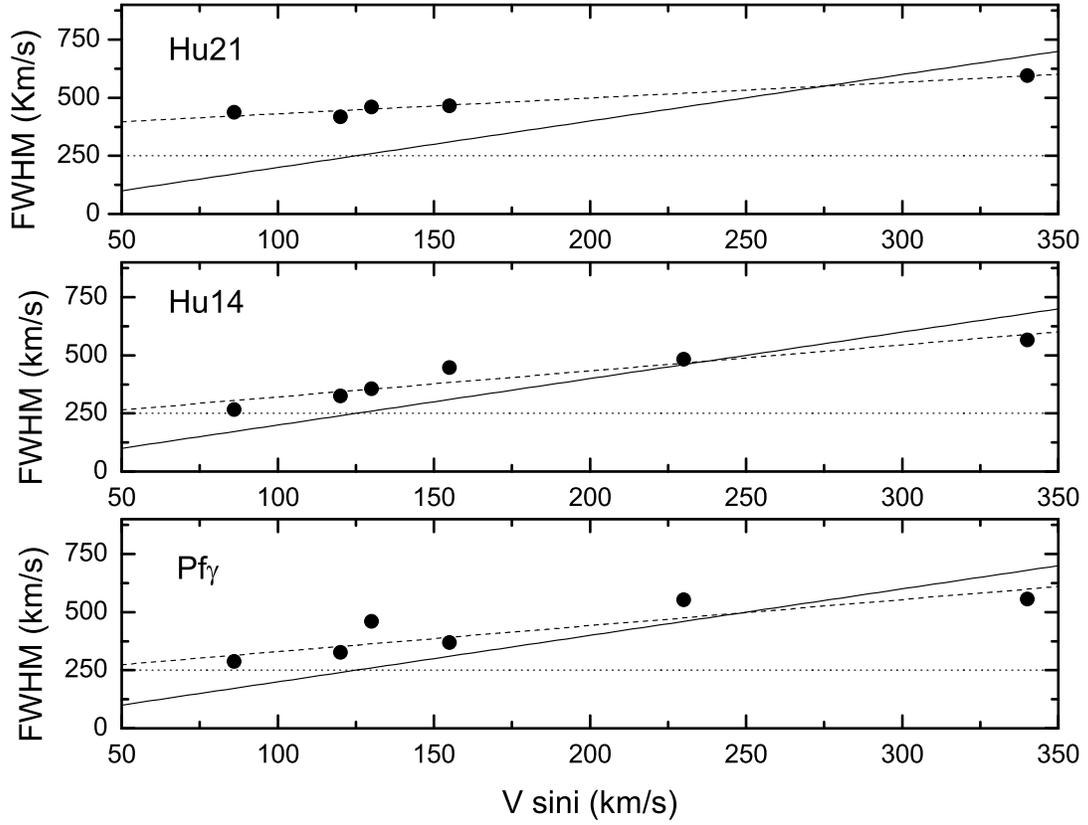}
\caption{$FWHM$ versus $V\,sin\,i$ for the program stars for lines Pf$\gamma$, Hu$_{14}$ and Hu$_{21}$. For each panel, the upper dashed line is the best linear fit, the solid line represents the case
$FWHM= 2\,V\,sin\,i$ and the dotted horizontal line indicates the spectral resolution (250 km/s).\label{f4}}
\end{figure}

\begin{figure}
\epsscale{1}
\plotone{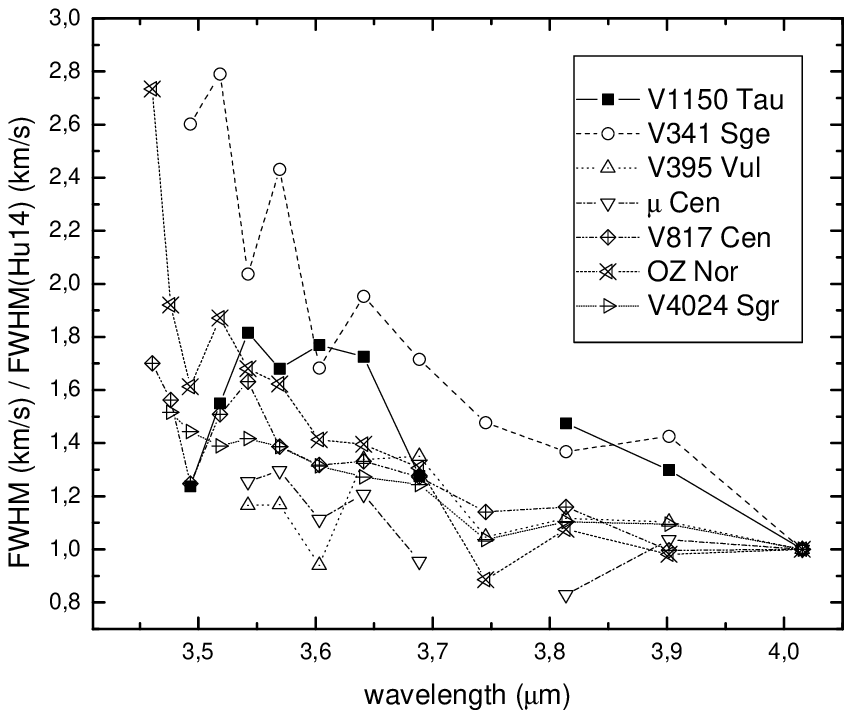}
\caption{The normalized full width at half maximum versus wavelength for the Humphreys emission lines.\label{f5}}
\end{figure}

\section{Discussion}

\subsection{Kinematical issues}

Our line width accuracy is limited by the instrumental resolution, the  signal to noise ratio of the spectra and the determination of the continuum level.
We estimate the error
of the full width at half maximum ($FWHM$) 
of the emission lines (corrected by the instrumental resolution and measured in the narrow slit spectra)
to be $\approx$ 40 km/s. On the other hand, the $V\,sin\,i$ values given in the literature are usually accurate
to less than $\sim$ 10\%, although for fast rotators the values could be underestimated by larger amounts (Townsend, Owocki \& Howarth 2004). Keeping in mind these caveats, we observe
that $FWHM$ roughly correlates with $V\,sin\,i$ and also is larger for higher order transitions in the Humphreys series  (Figs.\,4 and 5).  
We found correlations of the form (Table 4): \\

$FWHM=  A \times V\,sin\,i + B. $ \hfill(1)\\

\noindent  
 We considered only correlations with $R > 0.60$ and at most one rejected point.  
 These correlations suggest that rotational broadening is the main line broadening mechanism for these lines and point to a rotationally supported, probably disk-like envelope as the source of infrared line emission.
 Similar kinematical insights have been derived from other infrared and optical spectroscopic studies of Be stars and they have been interpreted in terms of  a disk-like geometry for Be star envelopes (e.g. Sellgren \& Smith 1992, Hanuschik 1996,
Clark \& Steele 2000, Hony et al. 2000). 
In this view, higher order lines probe inner disk regions, with larger rotational velocities.  The fact that the $FWHM$ is larger than 2\,$V\,sin\,i$ could indicate additional sources of broadening like turbulence, macroscopic velocity fields or electron scattering (although we do not observe prominent electron scattering wings in the lines).  We note that the $B$ coefficient defined in equation (1) is the expected line broadening for a star seen pole-on ($V\,sin\,i$ = 0 km/s). If  planar Keplerian motions dominate the disk kinematics, then we should expect this number to be similar to the thermal broadening in the disk ($B_{th}$ $\approx$ 13 km/s for 10.000 $K$ hydrogen gas).  However, we observe all lines in Table 4 with $B >>  B_{th}$.
Since IR lines, especially those of the Humphreys series, probe the inner disk region,  it is possible that larger turbulent motions and  eventually departures of the disk geometry in the inner disk explain why the $B$ coefficients of  infrared lines are larger than those of Balmer and Fe\,II lines (Table\,4). As we don't observe Stark broadened emission wings, we discard pressure effects  as the cause for  the large observed $B$. We failed to estimate the electron density of  the envelope (and so an estimate of the pressure effect)  using the Inglis-Teller formula since the quality of the data did not allow us to detect the wavelength of confluence  of the Humphreys series with the continuum.
The broadening effect mentioned, plus our rather small spectral resolution ($R \approx$ 1200), probably explains why  the typical hallmarks  for rotationally supported disks, viz.  doubly peaked emission lines, are not observed in our spectra. 

In the disk model of Be stars the peak separation $\Delta\lambda_{n}$ of the nth emission line measures the velocity $v_{n}$ near the outer disk (Hirata \& Kogure 1984):\\

$c\frac{\Delta \lambda_{n}}{\lambda_{n}}=v_{n}sin\,i$   \hfill(2)\\

\noindent
In the following we assume a disk rotational law given by:\\

$v(r)= v_{\star}(\frac{r}{r_{\star}})^{-j}$   $r \geq r_{\star}$  \hfill(3)\\

\noindent
where $v_{\star}$ is the equatorial stellar velocity and $j$ is equal to 0.5 for a Keplerian disk and 1.0
for continuum mass loss with conservation of the angular momentum. The extension of the disk which
corresponds to the nth emission line is:\\

$\frac{r_{Hun}}{r_{\star}}= (\frac{v_{Hun}}{v_{\star}})^{-1/j} $ \hfill(4) \\

 \noindent
 and the extension relative to the Hu14 forming region is:\\
 
 $\frac{r_{Hun}}{r_{Hu14}}= (\frac{v_{Hun}}{v_{Hu14}})^{-1/j}= (\frac{FWHM_{Hun}}{FWHM_{Hu14}})^{-1/j} $\hfill(5) \\
 
 \noindent
Where we have assumed a linear relation between $v_{Hun}$ and $FWHM_{Hun}$ (Hanuschik 1996 and references therein).
The fact that the ratio $(FWHM_{Hun}/FWHM_{Hu14})$ reaches values up to 2.7 around Hu24  (Fig.\,5), implies
that the relative disk extension $r_{Hu14}/r_{Hu24}$ for OZ Nor and V\,341 Sge equals 3 ($j$=1) or 7 ($j$=0.5). \\

\subsection{Diagnostics for the envelope optical depth}


\citet{blb} have suggested that the study of  $\log$ (Hu14/Br$\alpha$) versus $\log$ (Hu14/Pf$\gamma$)  provides a simple way to investigate the nature of the circumstellar material around Be stars and related objects. They found that line ratios close to the Menzel case B recombination theory correspond to optically thin lines formed mainly in isothermal stellar winds and those close to unity are optically thick lines arising from a disk-like structure. Particularly,  in an optically thick medium, the line flux ratios become independent on the  mass absorption coefficient and the line flux turns out to be dominated by the size of the emitting surface. We show in Fig.\,6 our decrements compared with those taken from the literature or from theoretical models. It is clear that Group II stars are distributed in a region of moderate or small optical depth, whereas those of Group I are in the extreme of the distribution, corresponding to cases of optically thick envelopes. Group III stars, without emission lines, probably correspond to stars that have lost most of their envelopes. We note that not all of these stars show Br$\alpha$ and Pf$\gamma$ absorption lines, suggesting the filling by emission in these lines and consecuently the presence of an incipient envelope. We find that our empirical groups trace the optical depth of the circumstellar envelope. 

We note the change in position of V\,395 Vul ($\equiv$ 12 Vul) in two epochs. 
As said in Section 3, this corresponds to a weakening of the H$\alpha$ emission line strength, which is
consistent with a more traslucent envelope in may 2003.
We also investigated the ISO spectra of the stars reported by \citet{blb} and published by Vandenbussche et al.  (2002). We note  
that above $\log$ (Hu14/Br$\alpha$) $\approx$ -0.2 all the stars (7 objects, including V\,1150 Tau) can be classified as Group I, and below that hypothetical line the stars (9 objects) are Group II.  12\,Vul changes between Group I and II. 
This shows that transition objects are difficult to find, and that the classification in groups I, II and III  is in principle observationally supported. This classification has a quantitative support in the \citet{blb} diagram, but can be done more rapidly by simple visual inspection of the spectrum, for classification or selection purposes. We note that Be stars showing long-lived outbursts are found only in the upper right part of the diagram whereas those showing short-lived outbursts are more widely distributed.  
Our qualitative conclusion about the optical depth condition in the envelopes of Group I \& II stars is also supported by the emission line ratio Pf$\gamma$/Hu16, which is a good discriminator between optically thin and thick conditions (Hummer \& Storey 1987, Hamann \& Simon 1987). We found this ratio $\sim$ 1 for all Group I stars and much larger than unity for Group II stars, consistent with theoretical predictions for optically thick and thin envelopes (respectively)
with $T$= 10.000 $K$ and $n_{e}$= 10$^{10}$ cm$^{-3}$.  

A quantitative interpretation of the Lenorzer et al. diagram was done by Jones et al. (2009),
in terms of a disk with varying density and illuminated by a central star of given temperature $T_{eff}$. In this model the disk density profile is given by:\\

$\rho(R,Z) = \rho_{0}(\frac{R_{\star}}{R})^{n}e^{-(Z/H)^2}$ \hfill(6) \\

\noindent
where $R$ is the cylindrical, radial distance from the star rotation axis, $Z$ is the perpendicular height above the equatorial plane and $H$ the corresponding scale height. $\rho_{0}$ is the density at the inner edge of the disk in the equatorial plane. Jones et al. (2009) explain the region occupied by Be stars in the Lenorzer et al. diagram by different values of $\rho_{0}$, $n$ and $T_{eff}$. 
They note that the models correctly explain the observed decrements if $n$ = 3.0 to 4.5 and $\rho_{0}$= 7.5$\times$ 10$^{-12}$ to 1.0 $\times$ 10$^{-11}$.  Their models do not reproduce, however, the extreme optically thick  cases located at the upper right corner of the diagram (our Group I stars). 
The fact that in Group I stars  Br$\alpha$ is comparable in strength to Humphreys lines suggests that the size of the emitting region is limited (by a putative companion for instance) and the gas density is high. However, the fact that we observe Br$\alpha$ line widths lower than in Humphreys lines for all Group I cases, argues against this view. Also, the Jones et al. (2009) disk models cannot reproduce these cases by simply truncating the disk since, as reported by these authors, all line fluxes fall off approximately at the same rate with increasing distance from the central star.
Jones et al. (2009) claim that these stars must have disk geometries different than a simple power-law. In this sense, different geometries like power-laws with non constant $n$  (Zorec et al 2007) could be explored. 
Our discovery of a significant number of Be stars in this region calls for additional theoretical work for explaining these cases.

\subsection{Study of the $EW$s and Humphreys emission line decrements}

We find  a strong correlation between $EW/\lambda$ and wavelength for a given series. This parameter increases with $\lambda$ and sometimes saturates at $EW/\lambda \approx  6 \times 10^{4}$ (Fig.\,7). This kind of behavior was observed in the prototype Be star  $\gamma$ Cas in several infrared H\,I series 
and interpreted in terms of a decrease in the line source  function in the outer parts of the circumstellar region (Hony et al. 2000).  Here we demonstrate that this is an usual behavior of Group-I and Group-II stars, being the pattern only disrupted by the larger emission found in the 
Br$\alpha$ and Pf$\gamma$ lines of Group-II stars. 

We have studied $f(Hu\,n)/f(Hu19)$ ($n$ is the quantum number of the upper level) versus $\lambda$ for every star showing emission lines.
We find that the Humphreys decrements follow well defined patterns (Fig.\,8). For the stars of Group II,  they increase with $\lambda$, indicating optically thin conditions, as in the case of $\beta$ Monocerotis A (Sellgren \& Smith 1992). On the contrary, for stars of Group I, the decrements change their behavior at a transition  wavelength, 
$\lambda_{0}$, from optically thin conditions (at shorter wavelengths) to optically thick conditions (at longer wavelengths). This transition is rather fast, and $\lambda_{0}$, given in Table 3, apparently is  a characteristic of every star. As in the case of  $EW/\lambda$, the transition at $\lambda_{0}$ could reflect a change in the optical properties of the envelopes at a certain distance of the star. In this scheme,
V\,1150 Tau could be classified as a Group-I star. 
We note that even for Be stars with optically thick envelopes,  high order Humphreys lines probe optically thin inner regions. 

From Fig.\,5 and 7 we deduce a rough anticorrelation between  $FWHM$ and $EW/\lambda$ for the Humphreys lines of a given star,
as usually happens for  the  H$\alpha$ line in Be stars (Dachs et al. 1986).  Consequently,
this can be explained assuming that the $EW$ scales with the disk size and that higher order lines 
are formed in smaller and inner disk-like
regions rotating faster than the outer and bigger regions forming low order Humphreys lines.

\begin{figure*}
\epsscale{1}
\plotone{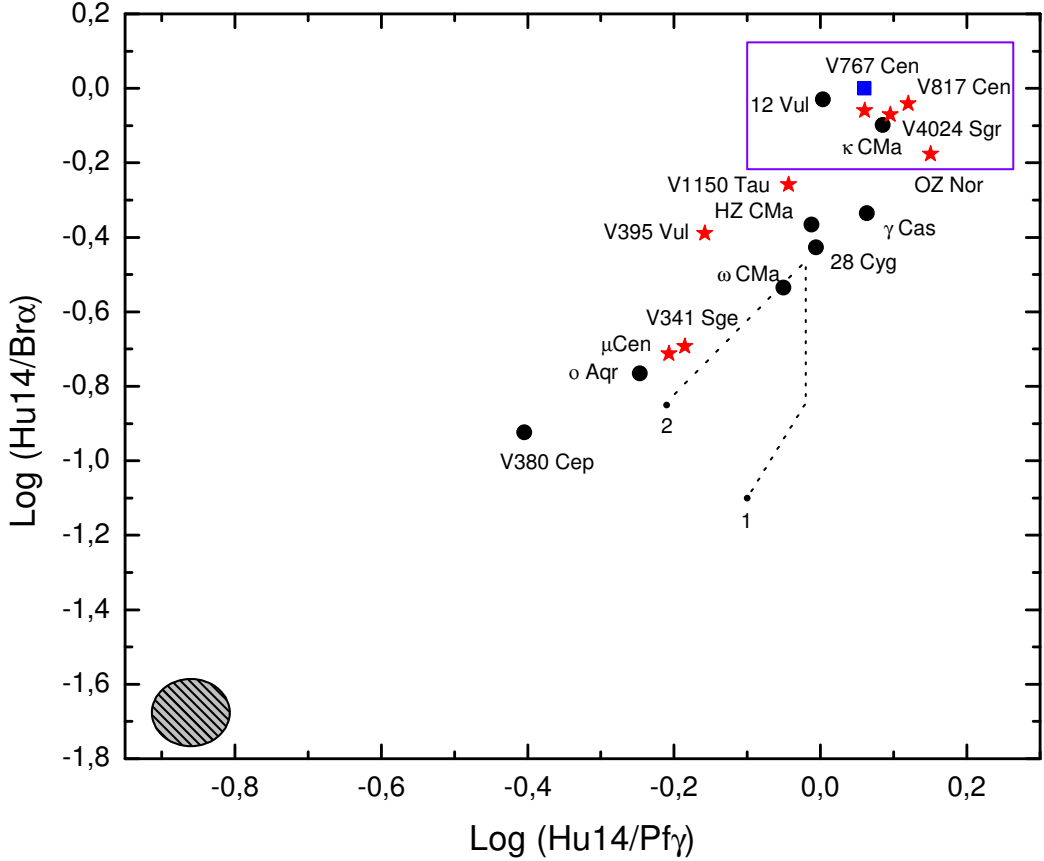}
\caption{Line ratio diagram  for our objects (stars) and Be stars from Lenorzer et al. (2002a) (dots). The solid square indicates the loci for an optically thick  black body and the big disk  the range of ratios for Menzel case B recombination, including collisional de-excitation, for temperatures higher than 10$^{4}$ $K$. Dashed lines illustrate the path of the models by Jones et al. (2009) for a 12 $M_{\sun}$ B2 star with luminosity 7000 $L_{\sun}$,  $T_{eff}$= 20.000 $K$, radius 7.0 $R_{\sun}$, log g= 4.0, inclination angle 35 degree and surrounded by a disk of solar composition, exponent $n$= 3.5 and increasing density $\rho_{0}$ from point 1 to 2. The model in the upper position has $\rho_{0}$= 10$^{-11}$ g cm$^{-3}$.  Above $\log$ (Hu14/Br$\alpha$) $\approx$ -0.2 all the stars  can be classified as Group I (stars inside the upper box), and below that hypothetical line the stars  are Group II. Note the change in position of V\,395 Vul (12 Vul). \label{f6}}
\end{figure*}


\begin{figure}
\epsscale{1}
\plotone{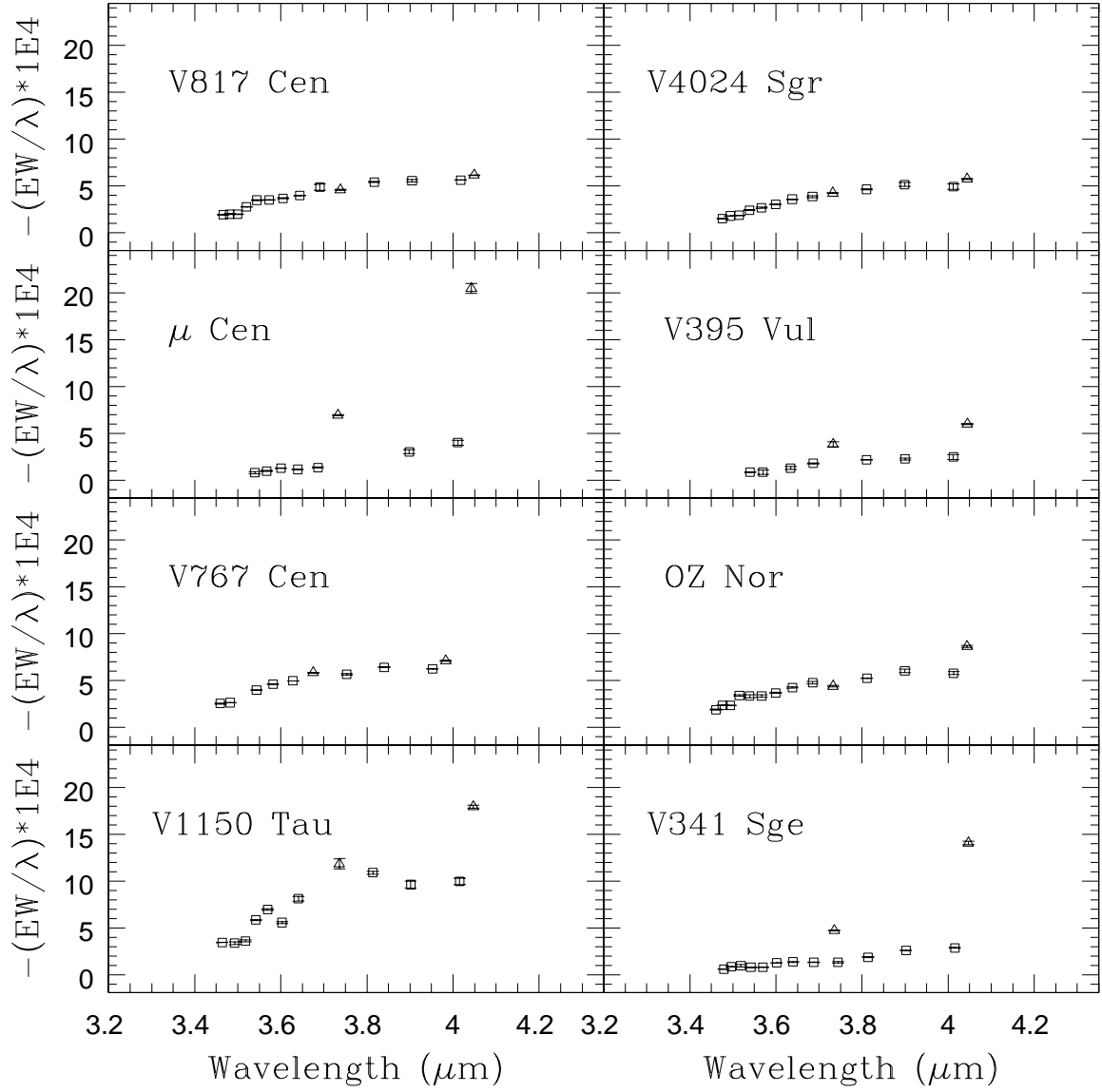}
\caption{$-EW/\lambda$ versus $\lambda$ for Br$\alpha$ and Pf$\gamma$ (triangles) and Humphreys (squares) lines for Be stars showing emission lines.\label{f7}}
\end{figure}

\begin{figure}
\epsscale{1}
\plotone{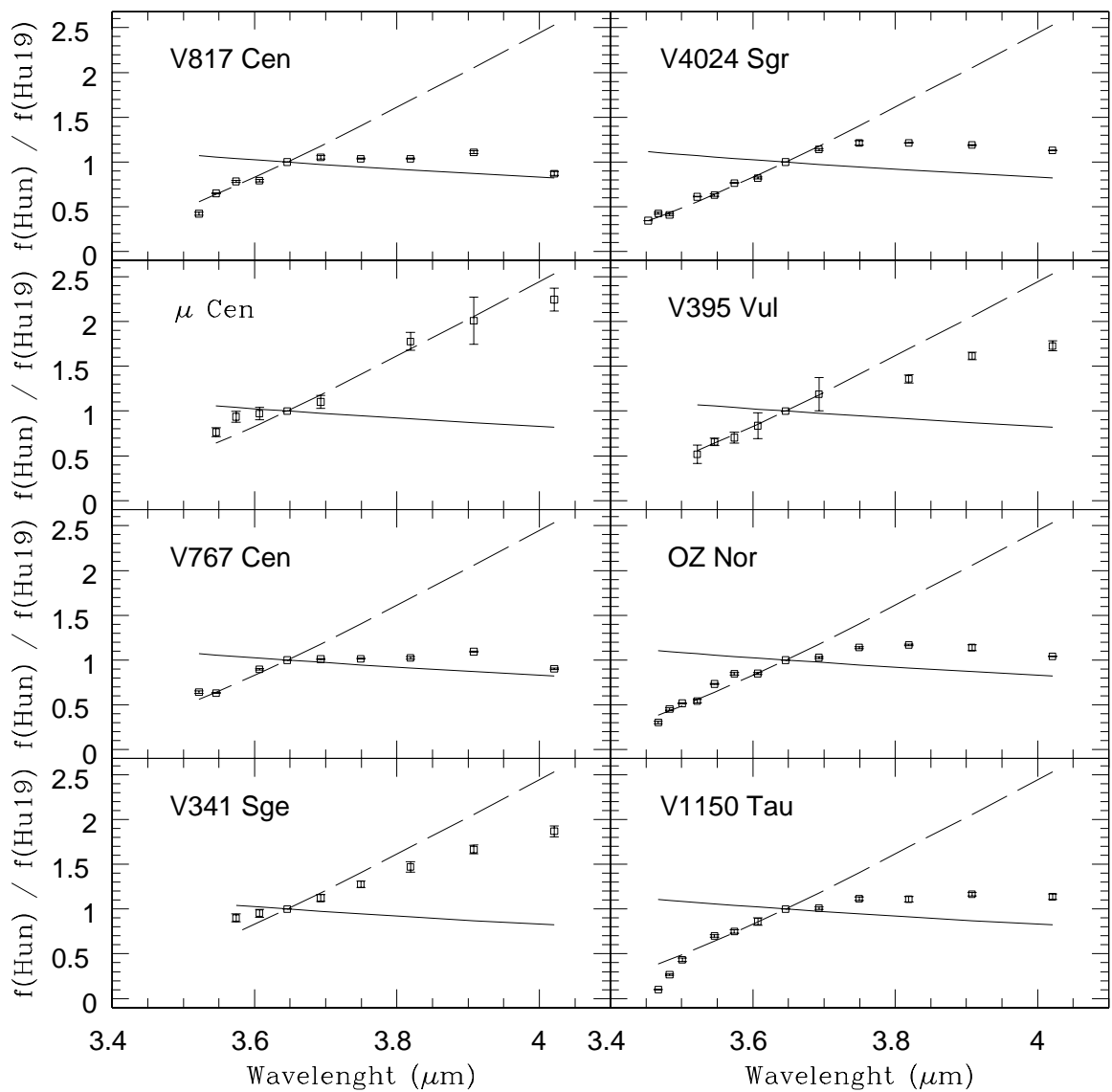}
\caption{$f(Hun)/f(Hu19)$ versus $\lambda$ for Be stars showing emission lines ($n$ is the quantum number of the upper level). The theoretical cases of optically thin Menzel Case-B recombination lines (dashed lines) and optically thick lines (solid lines) are also shown.\label{f8}}
\end{figure}

\subsection{2MASS photometry and continuum slope}

For each star we have calculated  $E(H-K) = (H-K)_{obs} - (H-K)_{0}$, where the first 
color is from 2MASS (Skrutskie et al. 2006) and the last one is  given by Koornneef (1983) for a star of similar spectral type and luminosity class (Table 2). We have not corrected them by interstellar extinction, which is expected to be  small in infrared wavelengths, especially for our bright and relatively close stars.   Keeping in mind that variability could affect the comparison of non-simultaneous data, we find no correlation between color excess, outburst character and group membership. 

We measured the flux ratio in regions almost depleted of emission lines:\\

flux ratio =  $\frac{\int_{3.41}^{3.47}S_{\lambda}d\lambda}{\int_{3.93}^{4.00}S_{\lambda}d\lambda}$ \hfill(7)\\

\noindent
where $S_{\lambda}$ is the spectral flux density. 
This ratio is bluer for hotter stars and for a given temperature is bluer for stars without emission lines (i.e. Group 3 stars, Fig.\,9). 
No correlation is observed between continuum slope and emission line strength. These findings possibly
point to the importance of the stellar flux in the infrared continuum emission; hotter stars produce bluer continuum
and when the disk becomes developed (Group I-II stars) this continuum becomes redder, probably due to the contribution of free-free emission 
and partial stellar obscuration from the optically thicker  disk. The position of 
V\,457 Sct is notable. This star shows the bluest continuum, and no developed emission. However, 
the absence of Br$\alpha$ and Pf$\gamma$ absorptions indicate that residual emission is filling these lines. Incidentally, V\,1448 Aql also show these characteristics; a very blue continuum and filled absorption lines. 
We speculate that these stars could be in a special position of their eruptive cycles, such as 
in the process of ejecting a hot optically thick  disk. The disk could be later dissipated as an optically thin ring
into the interstellar medium, as proposed in the model of de Wit et al. (2006), causing the stars' position to move down in the diagram. The bluer color during the outburst ascending branch are predicted by this model, and observed in Be outbursting stars (de Wit et al. 2006).  The stars should move up and down in the diagram of Fig.\,9 during the outburst rising and decay.  We note that Be star outbursts seem to be of larger amplitude in red bands (Mennickent et al. 2002) so this effect in 
our studied region should be significant. 
Further time resolved spectroscopy sampling the whole eruptive cycle is needed to 
test this conjecture for Be stars.


\begin{figure}
\epsscale{1}
\plotone{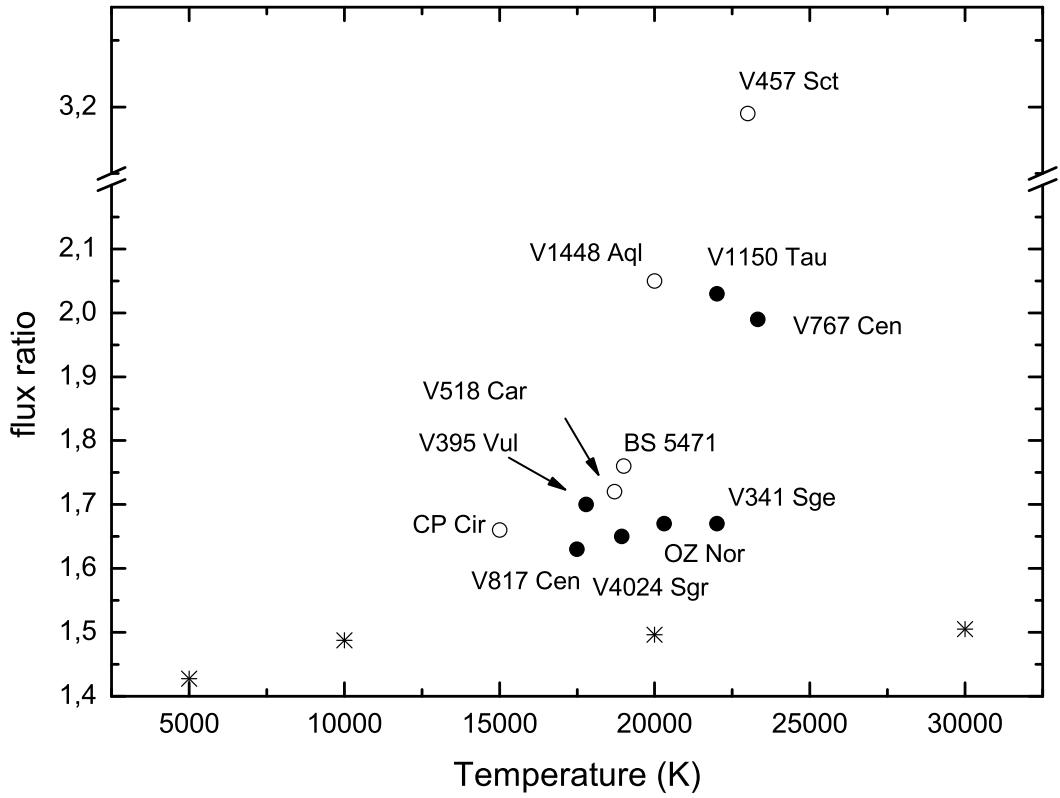}
\caption{The flux ratio defined in Eq.\,(7) versus the stellar effective temperature. Open circles indicate Group III stars and 
filled circles Group I and Group II stars. Asterisks show the position for simple black body radiators.\label{f9}}
\end{figure}

\subsection{Dust around outbursting Be stars?}

We note the absence of dust spectral features at the $L$ band, especially the Polycyclic Aromatic Hydrocarbons         
(PAH) emission feature at 
3.3  $\mu$m and the nano-diamond features at 3.43 and 3.52 $\mu$m, that have been detected in some Herbig Ae/Be stars. 
These features have been observed in pre-main sequence stars of spectral type B9 and later but for earlier spectral types the 3.3 $\mu$m band is weak or absent and the other bands are even weaker (Habart et al. 2004, Acke \& van den Ancker 2006). As our stars are mostly early B-type, we cannot establish the absence of dusty envelopes from the absence of these key spectral features. However,
 the infrared color excesses $E(H-K)$ of our targets listed in Table 2 and the 2MASS colors, $J-K$ between -0.08 and +0.34 and $H-K$ between -0.29 and 0.30,  are typical for Be stars  and not so large as in most hot stars surrounded by dust  (e.g. Fig.\, 11 in Mathew, Subramaniam \& Bhatt 2008).
 This suggests the absence of significant dust in the envelopes of these ourbursting Be stars. The  IRAS colors of our targets  provide the same insight.
This finding is consistent with the classification of our targets as "canonical" Be stars. Our targets  probably are not surrounded by massive cool envelopes as in the case of Herbig Ae/Be stars.

\section{Conclusions}

 

We have provided  a view of the L-band spectra of a selected sample of 
outbursting Be stars. 
These  spectra show no evidence of dust and are 
not different from those reported spectra of Be stars showing only irregular photometric variability.
The observed L-band spectra of 13 outbursting Be stars 
can be categorized in three broad groups reflecting the optical depth conditions in the Be star envelope. Based on the relative intensity of Humphreys, Br$\alpha$ and Pf$\gamma$ emission lines, 
 a rapid visual inspection of the spectra indicates the envelopes optical depth.  In addition, the Humphreys decrements, and the parameter $\lambda_{0}$ defined in Section 4.3, can be used as a diagnostic tool 
for the optical depth of the circumstellar envelopes. We find that higher order Humphreys lines probe optically thin inner regions even in the case of optically thick envelopes. In addition, the large broadening observed in the IR lines probably reflect vertical velocity fields near the star.  
Some power-law disk models describing the infrared emission line properties fail  to explain the cases of optically thick envelopes (Group I stars). We expect that our discovery of a large number of these stars motivates further theoretical work in this area. Our data does not allow us to test a possible correlation between the outburst stage and the spectral appearance,  but  the fact that the stars were observed at  random outburst phases,  the changes observed in two epochs in V\,395 Vul ($\equiv$ 12 Vul) and the blue continuums of V\,1448 Aql and V\,457 Sct, suggest that all outbursting stars observed in this project could pass  during their cycles through Groups I, II and III. 
The existence of these groups is in principle consistent with the proposed outburst description by de Wit et al. (2006) in terms of the ejection of an optically thick disk that expands and becomes optically thin before dissipation into the interstellar medium.
Accordingly, $\lambda_{0}$, and the whole spectral appareance  should change in a significant way during the entire outburst cycle, following the development  of the circumstellar envelope. Variability of Be stars in the diagram of Fig.\,6 along the diagonal was already suggested by Lenorzer et al. (2002b) due to the transient nature of Be star disks. We further suggest that the Be star outburst mechanism  and disk evolution could be tested with time resolved infrared spectroscopy, using for instance the spectral diagnostics worked in this paper.  Further time resolved L-band spectroscopy of selected targets along with photometry spanning several eruptive cycles are needed to test our prediction for outbursting Be stars.






\acknowledgments

We thank an anonymous referee for useful suggestions on a first version of this paper.
REM acknowledges financial support by Fondecyt grant 1070705, 
the Chilean Center for Astrophysics FONDAP 15010003 and from the BASAL
Centro de Astrof\'isica y Tecnologias Afines (CATA) PFB--06/2007.
BS acknowledges support by Programa de
Becas de Doctorado MECESUP UCO0209.
LC acknowledges the Agencia de 
Promoci\'on Cient\'{\i}fica y Tecnol\'ogica for supporting this research with the grant BID 1728 OC/AR PICT 03-12720  and PICT 111.

\clearpage


\begin{thebibliography}{}
\bibitem[\protect\citeauthoryear{Andrillat, Jaschek, \& Jaschek}{1988}]{b1} Andrillat Y., Jaschek M., Jaschek C., 1988, A\&AS, 72, 129  
\bibitem[\protect\citeauthoryear{Acke 
\& van den Ancker}{2006}]{2006A&A...457..171A} Acke B., van den Ancker M.~E., 2006, A\&A, 457, 171 
\bibitem[\protect\citeauthoryear{Bessell \& Brett}{1988}]{b3}
Bessell, M. S., \& Brett, J. M., 1988, PASP, 100, 1134
\bibitem[\protect\citeauthoryear{Briot}{1981}]{b4}
Briot D., 1981, A\&A, 103, 5 
\bibitem[\protect\citeauthoryear{Clark \& Steele}{2000}]{2000A&AS..141...65C} Clark J.~S., Steele I.~A., 2000, A\&AS, 141, 65 
\bibitem[\protect\citeauthoryear{Dachs et 
al.}{1986}]{1986A&A...159..276D} Dachs J., Hanuschik R., Kaiser D., Rohe D., 1986, A\&A, 159, 276 
\bibitem[\protect\citeauthoryear{de Wit et al.}{2006}]{b9} 
de Wit W.~J., Lamers H.~J.~G.~L.~M., Marquette J.~B., Beaulieu J.~P., 2006, 
A\&A, 456, 1027
\bibitem[\protect\citeauthoryear{Fr\'emat et al.}{2005}]{b10}
Fr\'emat Y., Zorec J., Hubert A.-M., Floquet M. 2005, A\&A, 440, 305
\bibitem[\protect\citeauthoryear{Fr\'emat et al.}{2006}]{b11}
Fr\'emat, Y.; Neiner, C.; Hubert, A.-M. et al. 2006, A\&A, 451, 1053
\bibitem[\protect\citeauthoryear{Gehrz, Hackwell \& Jones}{1974}]{b12}
Gehrz R. D., Hackwell  J. A., \& Jones T. W. 1974, ApJ, 191, 675
\bibitem[\protect\citeauthoryear{Glebocki \& Stawikowski}{2000}]{b13}
Glebocki, R., Stawikowski, A. 2000, AcA, 50, 509
\bibitem[\protect\citeauthoryear{Grundstrom 
\& Gies}{2006}]{2006ApJ...651L..53G} Grundstrom E.~D., Gies D.~R., 2006, ApJ, 651, L53 
\bibitem[\protect\citeauthoryear{Habart, Natta, 
\& Kr{\"u}gel}{2004}]{2004A&A...427..179H} Habart E., Natta A., Kr{\"u}gel E., 2004, A\&A, 427, 179 
\bibitem[\protect\citeauthoryear{Hamann 
\& Simon}{1987}]{1987ApJ...318..356H} Hamann F., Simon M., 1987, ApJ, 318, 356 
\bibitem[\protect\citeauthoryear{Hanuschik}{1989}]{1989Ap&SS.161...61H} Hanuschik R.~W., 1989, Ap\&SS, 161, 61 
\bibitem[\protect\citeauthoryear{Hanuschik}{1996}]{XXX} Hanuschik R.~W., 1996, A\&A, 308, 170 
\bibitem[\protect\citeauthoryear{Hirata 
\& Kogure}{1984}]{1984BASI...12..109H} Hirata R., Kogure T., 1984, BASI, 12, 109 
\bibitem[\protect\citeauthoryear{Hony et al.}{2000}]{b14}
Hony S. et al. 2000, A\&A, 355, 187
\bibitem[\protect\citeauthoryear{Hubert, Floquet \& Zorec}{2000}]{b16}
Hubert A. M., Floquet M. \& Zorec  J., 2000, in IAU Colloquium 175, `` The Be Phenomenon in Early-Type Stars'', Eds. M. A. Smith and H. F. Henrichs, ASP ConferenceProceedings, Vol. 214, p. 348
\bibitem[\protect\citeauthoryear{Hummer 
\& Storey}{1987}]{1987MNRAS.224..801H} Hummer D.~G., Storey P.~J., 1987, MNRAS, 224, 801 
\bibitem[\protect\citeauthoryear{Jaschek \& Jaschek}{1987}]{b17} 
Jaschek C., \& Jaschek M. 1987, "The Classification of Stars", Cambridge University Press.
\bibitem[\protect\citeauthoryear{Jones et al.}{2009}]{2009MNRAS.392..383J} 
Jones C.~E., Molak A., Sigut T.~A.~A., de Koter A., Lenorzer A., Popa 
S.~C., 2009, MNRAS, 392, 383 
\bibitem[\protect\citeauthoryear{Koornneef}{1983}]{1983A&A...128...84K} Koornneef J., 1983, A\&A, 128, 84 
\bibitem[\protect\citeauthoryear{Lenorzer et 
al.}{2002a}]{bla} Lenorzer A., Vandenbussche B., Morris P., de Koter A., Geballe T.~R., Waters L.~B.~F.~M., Hony S., Kaper L., 2002a, A\&A, 384, 473 
\bibitem[\protect\citeauthoryear{Lenorzer, de Koter, 
\& Waters}{2002b}]{blb} Lenorzer A., de Koter A., Waters L.~B.~F.~M., 2002b, A\&A, 386, L5 
\bibitem[\protect\citeauthoryear{Maiolino, Rieke, 
\& Rieke}{1996}]{1996AJ....111..537M} Maiolino R., Rieke G.~H., Rieke M.~J., 1996, AJ, 111, 537 
\bibitem[\protect\citeauthoryear{Mathew, Subramaniam, 
\& Bhatt}{2008}]{2008MNRAS.388.1879M} Mathew B., Subramaniam A., Bhatt B.~C., 2008, MNRAS, 388, 1879 
\bibitem[\protect\citeauthoryear{Mennickent, Vogt, 
\& Sterken}{1994}]{1994A&AS..108..237M} Mennickent R.~E., Vogt N., Sterken C., 1994, A\&AS, 108, 237
\bibitem[\protect\citeauthoryear{Mennickent et al.}{2002}]{b27} 
Mennickent R.~E., Pietrzy{\'n}ski G., Gieren W., Szewczyk O., 2002, A\&A, 393, 887 
\bibitem[\protect\citeauthoryear{Moorwood et 
al.}{1998}]{1998Msngr..94....7M} Moorwood A., et al., 1998, Msngr, 94, 7 
\bibitem[\protect\citeauthoryear{Persson \& McGregor}{1985}]{b30}
Persson S. E. \& McGregor P. J. 1985, AJ, 90, 1860
\bibitem[\protect\citeauthoryear{Pojma{\'n}ski}{2001}]{p1} 
Pojma{\'n}ski G., 2001, ASPC, 246, 53 
\bibitem[\protect\citeauthoryear{Quirrenbach et 
al.}{1994}]{1994A&A...283L..13Q} Quirrenbach A., Buscher D.~F., Mozurkewich D., Hummel C.~A., Armstrong J.~T., 1994, A\&A, 283, L13 
\bibitem[\protect\citeauthoryear{Sellgren 
\& Smith}{1992}]{1992ApJ...388..178S} Sellgren K., Smith R.~G., 1992, ApJ, 388, 178 
\bibitem[\protect\citeauthoryear{Skrutskie et 
al.}{2006}]{2006AJ....131.1163S} Skrutskie M.~F., et al., 2006, AJ, 131, 
1163 
\bibitem[\protect\citeauthoryear{Stee et 
al.}{1995}]{1995A&A...300..219S} Stee P., de Araujo F.~X., Vakili F., Mourard D., Arnold L., Bonneau D., Morand F., Tallon-Bosc I., 1995, A\&A, 300, 219 
\bibitem[\protect\citeauthoryear{Sterken, Vogt, 
\& Mennickent}{1996}]{1996A&A...311..579S} Sterken C., Vogt N., Mennickent R.~E., 1996, A\&A, 311, 579 
\bibitem[\protect\citeauthoryear{Townsend, Owocki, 
\& Howarth}{2004}]{2004MNRAS.350..189T} Townsend R.~H.~D., Owocki S.~P., Howarth I.~D., 2004, MNRAS, 350, 189 
\bibitem[\protect\citeauthoryear{Vandenbussche et 
al.}{2002}]{2002A&A...390.1033V} Vandenbussche B., et al., 2002, A\&A, 390, 1033 
\bibitem[\protect\citeauthoryear{Waters}{1986}]{b45}
Waters L. B. F. M. 1986, A\&A, 162, 121 
\bibitem[\protect\citeauthoryear{Wright et al.}{2003}]{b47}
Wright C. O., Egan M. P., Kraemer K. E., Price S. D. 2003, AJ, 125,359
\bibitem[\protect\citeauthoryear{Yudin}{2001}]{b48}
Yudin R. V. 2001, A\&A, 368, 912
\bibitem[\protect\citeauthoryear{Zorec, Fr\'emat \& Cidale}{2005}]{b49}
Zorec J., Fr\'emat Y. \& Cidale L. 2005, A\&A, 441, 235
\bibitem[\protect\citeauthoryear{Zorec et 
al.}{2007}]{2007A&A...470..239Z} Zorec J., Arias M.~L., Cidale L., Ringuelet A.~E., 2007, A\&A, 470, 239 

\end{thebibliography}
\end{document}